\documentclass[aps, pra, showpacs, superscriptaddress, twocolumn, 10pt]{revtex4-1}
\usepackage{graphicx}
\usepackage{dcolumn}
\usepackage{bm}
\usepackage{amsmath}
\usepackage{euscript}
\usepackage[percent]{overpic}
\usepackage{subfig}
\usepackage{amsfonts,amssymb,amsmath,tabularx,booktabs}
\usepackage{graphics,graphicx,color,enumitem}
\usepackage{epstopdf,epsfig}
\usepackage{multirow,array,varwidth,spreadtab,caption}
\usepackage{siunitx,booktabs,caption}
\usepackage[colorlinks=true,linkcolor=blue,citecolor=blue]{hyperref}

\begin{document}
\newcommand{\ket}[1]{\left | #1 \right\rangle}
\newcommand{\bra}[1]{\left \langle #1 \right |}
\setlength{\tabcolsep}{4pt}
\renewcommand{\arraystretch}{1.2}

\title{Survey on the Bell nonlocality of a pair of entangled qudits}

\author{Alejandro Fonseca}
\affiliation{Departamento de F\'{i}sica, Universidade Federal de Pernambuco, 50670-901, Recife, Pernambuco, Brazil}
\author{Anna de Rosier}
\affiliation{Institute of Theoretical Physics and Astrophysics, Faculty of Mathematics, Physics and Informatics, University of Gda\'nsk, 80-308 Gda\'nsk, Poland}
\author{Tam\'as~V\'ertesi}
\affiliation{Institute for Nuclear Research, Hungarian Academy of Sciences, H-4001 Debrecen, P.O. Box 51, Hungary}
\author{Wies{\l}aw~Laskowski}
\affiliation{Institute of Theoretical Physics and Astrophysics, Faculty of Mathematics, Physics and Informatics, University of Gda\'nsk, 80-308 Gda\'nsk, Poland}
\author{Fernando Parisio}
\affiliation{Departamento de F\'{i}sica, Universidade Federal de Pernambuco, 50670-901, Recife, Pernambuco, Brazil}


\begin{abstract}
The question of how Bell nonlocality behaves in bipartite systems of higher dimensions is addressed. By employing the probability of violation of local realism under random measurements as the figure of merit, we investigate the nonlocality of entangled qudits with dimensions ranging from $d=2$ to $d=7$. We proceed in two complementary directions. First, we study the specific Bell scenario defined by the Collins-Gisin-Linden-Massar-Popescu (CGLMP) inequality. Second, we consider the nonlocality of the same states under a more general perspective, by directly addressing the space of joint probabilities (computing the frequencies of behaviours outside the local polytope). In both approaches we find that the nonlocality decreases as the dimension $d$ grows, but in quite distinct ways. While the drop in the probability of violation is exponential in the CGLMP scenario, it presents, at most, a linear decay in the space of behaviours. Furthermore, in both cases the states that produce maximal numeric violations in the CGLMP inequality present low probabilities of violation in comparison with maximally entangled states, so, no anomaly is observed. Finally, the nonlocality of states with non-maximal Schmidt rank is investigated.

\end{abstract}

\maketitle

\section{\label{I}Introduction}
The violation of Bell inequalities \cite{Bell64}, recently confirmed by experiments not afflicted by detection and locality loopholes \cite{loophole1,loophole2,loophole3,PhysRevLett.119.010402}, constitutes one of the most impressive confirmations of the nonlocal character of quantum theory. Presently, the majority of the state-of-the-art experiments in the field involve two qubits in the 
context of the Clauser-Horne-Shimony-Holt (CHSH) inequality. However, it became clear that the use of systems of higher dimensionality, or \textit{qudits}, may lead to
new, interesting phenomena and improvements in the efficiency of some practical tasks \cite{Mischuck13,Strauch11,Lanyon09,Ralph07,Durt04}. 
In particular, it may be easier in the future to carry out loophole-free Bell tests if qudits are employed \cite{vertesi}. 
The nonlocality of pairs of entangled qudits have been used to certify high dimensional entanglement and
in the study of robustness against noise, imperfect state preparation and measurements \cite{Dada11,Weiss16,Dutta16,Polozova16}.
Apart from its foundational relevance, Bell nonlocality is a primary resource within the field of quantum information \cite{Brunner14,Buhrman10}.

A more specific, but important question refers to the macroscopic limit. 
Pioneering works, addressing two spin-$s$ particles, revealed a tendency toward local, classical behaviours as 
$s \rightarrow \infty$ \cite{Mermin80,Mermin82}, in the sense that the range of parameters for which nonclassicality arises vanishes as 
$1/s$ (however the considered inequalities are not tight).
Complementarily, Gisin and Peres \cite{Gisin92} showed that, for particular choices of measurement parameters
in the context of the CHSH inequality, it is always possible to obtain violations, but not above the Tsirelson bound. 

The authors of \cite{Kaszlikowski00} employed the resistance to noise as a nonlocality quantifier, and numerically calculated it for maximally entangled states of two qudits up to $d=9$, each subject to one out of two local measurements characterized by 
multiport beam splitters and phase shifters (MBSPS) \cite{Zukowski97}. Rather surprisingly, the authors found that the resistance to white noise increases with the dimension $d$. Presently, it is acknowledged that, although physically relevant, resistance to noise is not a good measure of nonlocality. 
Also in this context, a surprising result is that the nonlocality of a system of $n$ qubits tends to increase with $n$, provided that the ability to individually address each qubit is preserved \cite{us}.

Further results indicated that the states that maximally violate the  Collins-Gisin-Linden-Massar-Popescu (CGLMP) inequality \cite{CGLMP} do not correspond to maximally entangled states for $d>2$ \cite{Acin02} (this is also valid for optimal Bell tests \cite{Zohren08,Acin05}). This unexpected finding has been considered as an ``anomaly'' of nonlocality.
In this context the probability of violation under random measurements \cite{Liang10,Wallman11} has been proposed as a measure of nonlocality \cite{Fonseca15}, and, contrary to these previous works, led to the conclusion that maximally entangled qutrits are maximally nonlocal. This indicates that the anomaly \cite{Methot07} in the nonlocality of entangled qudits may be an artefact of the previously employed measures (see however \cite{camalet}). Recently, other promising quantifiers have been proposed, as, for example, a trace distance measure (within the context of a resource theory for nonlocality) \cite{PhysRevA.97.022111}, and a nonanomalous realism-based measure \cite{PhysRevA.97.012123,arxivAngelo}.

In this work we employ the probability of violation to quantify the nonlocality of two entangled qudits up to $d=7$, in two distinct, complementary perspectives. First, we address a specific experimental situation, i. e., a fixed Bell scenario (CGLMP) and the set of observables which are accessible in a particular experimental realization, namely, MBSPS. Second, we investigate the same set of states in a more fundamental perspective, by calculating the probability of violation directly in the full space of joint probabilities (the space of behaviours). While the first approach corresponds to a situation that can be exhaustively investigated within a single experimental preparation, it also inherits the bias associated with the choice of a particular facet of the local polytope. 
The second approach is conceptually more powerful, since it takes into account all possible Bell inequalities (with a certain number of observables per party), however, the probabilities of violation calculated in the space of behaviours cannot possibly be determined by a single experimental setup. We discuss, both the common points and the differences between the two approaches.
\section{\label{II} Nonlocality of two entangled qudits in the CGLMP scenario}
We start by relating the volume of violation, defined as a quantifier of Bell nonlocality in \cite{Fonseca15}, with the probability of violation under random, directionally unbiased measurements. Here, the nonlocality extent of a quantum state $\rho$ within the scenario of a particular Bell inequality $I$ will be associated with:
\begin{equation}
 V_I(\rho) \equiv \frac{1}{\mathcal{N}} \int_{\Gamma_{\rho}} d^{n}x, 
\label{vol}
\end{equation}
where $\mathcal{X}=\{x_i\}$ is the set of all parameters that characterizes the measurements, 
$\Gamma_{\rho}\subset\mathcal{X}$ is the subset of parameters that lead to violation of the Bell inequality and $\mathcal{N}$ is a normalization constant.
In order to obtain the probability of violation, $p_{v}(\rho)$, we must write
\begin{displaymath}
\frac{1}{\mathcal{N}} =\frac{\nu}{V_{\mathcal{X}}} ,
\end{displaymath}
where $V_{\mathcal{X}}$ gives the total volume of the set of measurement parameters, 
\begin{displaymath}
 V_{\mathcal{X}} \equiv \int_{\mathcal{X}} d^{n}x, 
\end{displaymath}
and $\nu$ is the number of ways one can relabel Alice's and Bob's observables (the symmetry between Alice and Bob themselves, is already considered).
Since we have two observables per party in both the CHSH and CGLMP scenarios, throughout this work, $\nu=4$. 
In this way, $V_{I}(\rho) \rightarrow p_{v}(\rho)$ becomes a probability, which is the quantity that we will consider hereafter.
A complementary approach was recently used by Atkin and Zohren \cite{Atkin15},
in which the measurement settings are fixed and the number of outcomes of the
measurements is varied for several ensembles of random pure states. 
\subsection{Multiport beam splitters and phase shifters}
We will be concerned with bipartite systems with Alice and Bob sharing a pure entangled state $|\Psi \rangle$ of two $d$-level systems.
The state of such a system can always be written as a Schmidt decomposition: 
\begin{equation}
\label{State1}
 |\Psi \rangle = \sum_{j=0}^{d-1} \alpha_j |j \rangle_A \otimes |j \rangle_B.
\end{equation}
\begin{figure}
  \includegraphics[scale=0.4]{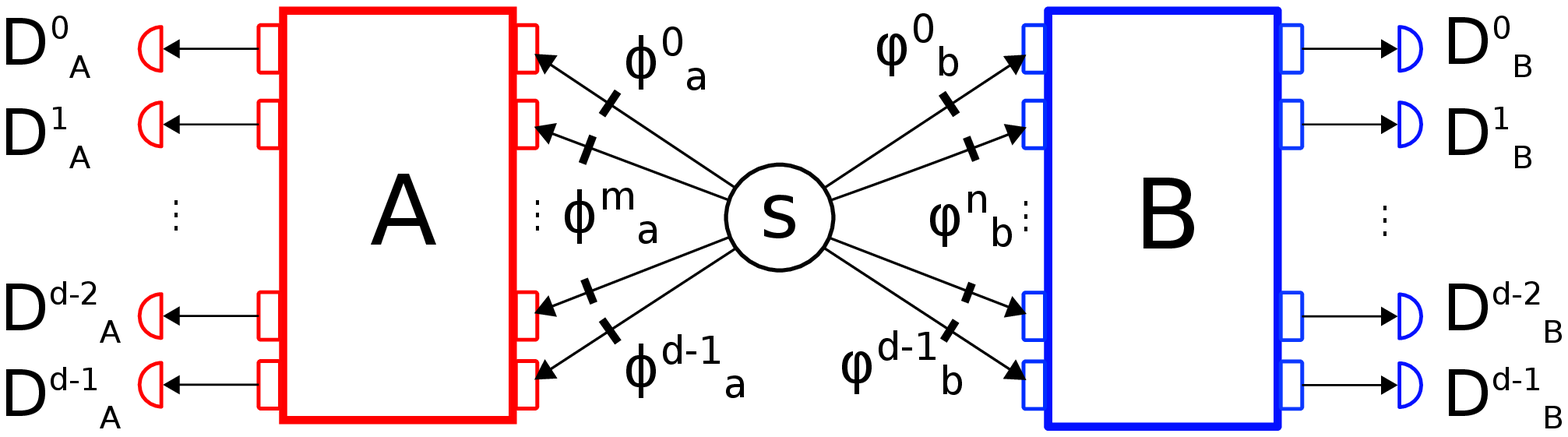}%
 \caption{\label{Fig1}(color online) Schematic illustration of the Multiport-Beam-Splitters-and-Phase-Shifters (MBSPS) realization of the CGLMP inequality.}
 \end{figure}
Each of the parties can execute one out of two $d$-outcome projective measurements ($a,b=1,2$) limited to a MBSPS scheme, which consists in 
diagonal phase-shift unitary operations: $U_{mm}=e^{i\phi^m_a}$ (Alice) and $U_{nn}=e^{i\varphi^n_b}$ (Bob), followed by  
discrete Fourier transforms $U_{FT}$ and $U_{FT}^*$ on Alice's and Bob's subsystems, respectively, and then a projection onto the original basis 
\cite{Kaszlikowski00,Zukowski97,Durt01,CGLMP,Zohren08,Acin05} (see fig. \ref{Fig1}). It is important to note that this doesn't exhaust the CGLMP scenario, however, 
we obtain a great simplification by remaining within MBSPS realizations, which are often employed in CGLMP-tests. In addition, this was exactly the considered situation when the anomaly in the nonlocality of two qutrits was first reported. It has also been conjectured that the optimal settings are contained in the MBSPS scenario \cite{Durt01}, which has been proved in the two-qutrit case in \cite{PhysRevLett.113.040401}.
    
The joint probability associated with the $k$-th and $l$-th outputs for Alice and Bob, respectively, given that 
their choices of observable were $a$ and $b$ reads:
\begin{equation}
\label{PQM}
P_{ab}(k,l)=\frac{1}{d^2}+\frac{2}{d^2}\sum_{m>n=0}^{d-1} \Re(\alpha_m\alpha^*_{n}) \cos\Delta^{mn}_{ab}(k,l),
\end{equation}
with
$$\Delta^{mn}_{ab}(k,l)=\phi^m_a+\varphi^m_b-\phi^{n}_a-\varphi^{n}_b+\frac{2\pi}{d}(m-n)\big(k\oplus(-l)\big),$$
where $\oplus$ denotes sum modulo $d$.

The corresponding CGLMP inequality is a facet of the associated local polytope \cite{Masanes03} and reads:
\begin{equation}
\label{ECGLMP}
I_d = \sum_{k=0}^{[d/2]-1}\left(1-\frac{2k}{d-1}\right)\Big\{\mathcal{B}_k-\mathcal{B}_{-(k+1)}\Big\}\leq 2, 
\end{equation}
here $[x]$ indicates the integer part of $x$ and $\mathcal{B}_k=P(A_1=B_1+k)+P(B_1=A_2+k+1)+P(A_2=B_2+k)+P(B_2=A_1+k)$, where
$P(A_a=B_b+k)$ is the probability that the outcomes corresponding to the observables $A_a$ and $B_b$ differ by $k$, modulo $d$.

Introducing the joint probabilities (\ref{PQM}) into (\ref{ECGLMP}) the CGLMP-Bell function can be rewritten in a simpler form, compatible with the MBSPS constraints (see the Appendix):
\begin{equation*}
I_d=\sum_{a,b=1}^{2}\sum_{m>n=0}^{d-1}C_{ab}^{mn}\cos(\phi^m_a+\varphi^m_b-\phi^{n}_a-\varphi^{n}_b+\varPsi_{ab}^{mn}),  
\end{equation*}
with coefficients $C_{ab}^{mn}$ and $\varPsi_{ab}^{mn}$ given by (\ref{AmpAP}) and (\ref{PhCGLMP}). 

The volume element of the set of measurement parameters is simply given by $d\Phi=\prod_{a,b=1}^{2}\prod_{j,k=0}^{d-1}d\phi^{j}_a d\varphi^{k}_b$. This ``trivial'' measure is due to the fact that all involved parameters are in-plane angles (in the MBSPS scheme). The total volume 
is $V_{\mathcal{X}}=(2\pi)^{4d}$, then the probability of violation may be calculated as:
\begin{equation}
p_{v}(\rho)=\frac{4}{(2\pi)^{4d}} \int_{\Gamma_{\rho}} d\Phi ,
\end{equation}
where $\Gamma_{\rho}$ corresponds to the subset of $\mathcal{X}$ for which the measurement parameters lead to violation of the inequality $I_d\leq 2$ for a given state 
$\rho$.

The results presented in this section have been obtained via Monte Carlo integrations, corresponding to several runs of a Bell experiment using uniform random measurement configurations on a definite quantum state.

Calculations of the probability of violation of pairs of qudits in maximally entangled states (MES) and maximally violating states (MVS) under the CGLMP inequality and MBSPS measurements were carried out up to $d=7$. The results are shown in a monolog plot in figure \ref{Fig2}. As it can be seen, the higher the dimension, the lower the probability of violation. In this way it is possible to conclude that the nonlocal content of a quantum entangled state of two qudits {\it exponentially} decreases with the dimensionality of the system, which is in agreement with the notion of restoration of classical features in the limit of high quantum numbers. However, we stress that the CGLMP scenario refers to two observables per party, no matter the value of $d$. We found that the exponential-decay behaviour assumes a particularly simple form if we use $2\pi$ as the basis (this is a natural basis in MBSPS scenarios). The points are well described by
\begin{equation}
p_v(d)\sim(2\pi)^{-d},
\end{equation}
where $p_v(d)$ refers to the maximally entangled state (MES) of two qudits with $d$ levels each. In figure \ref{Fig2}, these points are represented by (red) triangles, and the upper continuous line corresponds to the best fitting with $p_v(d)\sim(2\pi)^{-1.04d}$. The squares correspond to the states that yield the maximal numeric violation of the CGLMP inequality. Except for $d=2$ (for which equal probabilities are obtained), the MES
present a higher probability in comparison with the maximally violating states. The probability of violation for the MVS's drops off approximately as $(2\pi)^{-1.07d}$.
This extends the conclusion of \cite{Fonseca15}, showing that there is no anomaly in the nonlocality of two entangled qudits up to $d=7$, at least in the CGLMP scenario, when $p_v$ is used as a figure of merit.  Below, we provide a list of numerically calculated MVS's for $3\le d \le7$:
\begin{figure}
  \includegraphics[scale=0.5]{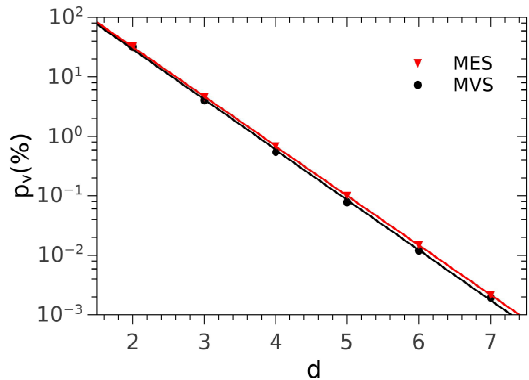}%
 \caption{\label{Fig2}(color online) Monolog plot of the probabilities of violation (in percents) of the maximally entangled state (MES) and maximally violating state (MVS) as a function of the dimension $d$ (CGLMP inequality and MBSPS measurements). The nonlocality decreases exponentially with the dimension.
 Note that apart from the qubit case ($d=2$), the MES presents more nonlocality than the MVS.}
 \end{figure}
\begin{widetext}
\begin{eqnarray}
	\ket{\psi_{\rm{MVS}}^{\rm{rank}=3}}&=&0.6169\ket{00}+0.4888\ket{11}+0.6169\ket{22},\\
	\ket{\psi_{\rm{MVS}}^{\rm{rank}=4}}&=&0.5686\ket{00}+0.4204\ket{11}+0.4204\ket{22}+0.5686\ket{33},\\
	\ket{\psi_{\rm{MVS}}^{\rm{rank}=5}}&=&0.5368\ket{00}+0.3859\ket{11}+0.3859\ket{22}+0.3859\ket{33}+0.5368\ket{44},\\
	\ket{\psi_{\rm{MVS}}^{\rm{rank}=6}}&=&0.5137\ket{00}+0.3644\ket{11}+0.3214\ket{22}+0.3214\ket{33}+0.3644\ket{44}+0.5137\ket{55},\\
	\ket{\psi_{\rm{MVS}}^{\rm{rank}=7}}&=&0.4957\ket{00}+0.3493\ket{11}+0.3011\ket{22}+0.2882\ket{33}+0.3011\ket{44}+0.3493\ket{55}+0.4957\ket{66}.
	\end{eqnarray}
\end{widetext}

The first three states coincide with those calculated in \cite{Zohren08}.
The MES and MVS coincide for $d=2$, and $p_v(2)\approx 0.32$, which shows that the restriction to MBSPS measurements increases the probability of violation. For general measurements, the probability of violation is around 0.28 for maximally entangled states, since the CGLMP and the CHSH inequalities are equivalent for $d=2$. A similar result appears when, in the CHSH scenario, the parties previously agree on one of the measurement directions. With this the inequality becomes the first Bell inequality, for which $p_v=1/3\approx 0.33$ \cite{parisio16}.

Regarding two qudits, MES are also maximally symmetric. However, one can consider maximally symmetric states (MSS) with Schmidt ranks such that $r<d$, which are not maximally entangled. In this case, the inequivalence between MSS's and states that maximize $p_v$ reappears for the CGLMP inequality.
In spite of the  balancedness of  states like $(|00\rangle+|11\rangle+ \cdots + |(r-1)(r-1)\rangle)/\sqrt{r}$, due to the fact
that the basis kets $|rr\rangle, \cdots , |dd\rangle$ are missing, they are not maximally nonlocal, in the CGLMP scenario. However, this doesn't constitute a true anomaly, since the symmetric low rank states cannot be considered maximally entangled. The investigation of states with lower ranks will provide a clear illustration of how different can the results be when a single Bell inequality is considered instead of the full space of behaviours.

As an example, let us consider the family of states (with zero as the coefficient of $\ket{33}$):
\begin{equation}
\label{Par}
\cos\theta_0 |00\rangle+\sin \theta_0 \cos\theta_1 |11\rangle+\sin \theta_0\sin \theta_1 |22\rangle.
\end{equation}
\begin{figure}
	\centering
	\includegraphics[scale=0.3]{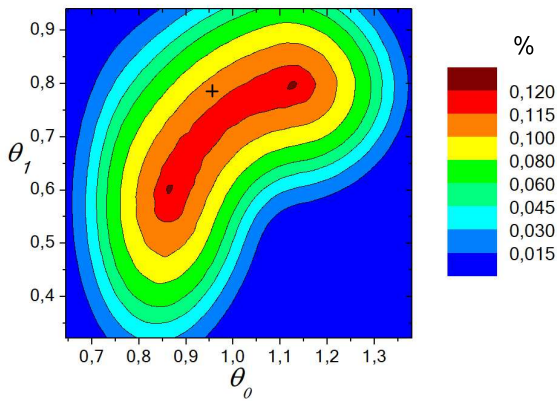}
	\caption{(color online) Probability of violation ($\%$) for rank-3 states with $d=4$ in the context of the CGLMP inequality. The cross corresponds to the state
	$(|00\rangle+|11\rangle+|22\rangle)/\sqrt{3}$, and the lower-left darker spot corresponds to state (\ref{rank3}).}
	\label{G437}
\end{figure}
In Fig. \ref{G437}
we plot $p_v$ for the above rank-3 states with $d=4$,  as a function of $\theta_0$ and $\theta_1$. The balanced state is identified by the cross,
while the two states that maximize the probability of violation are given by $(\theta_0,\theta_1)\approx(0.864,0.604)$, 
\begin{eqnarray}
\label{rank3}
\nonumber
0.647\ket{00}+0.628\ket{11}+0.431\ket{22};
\end{eqnarray}
and $(\theta_0,\theta_1)\approx(1.126,0.798)$ (equivalent to the above state with $\ket{00} \leftrightarrow \ket{22}$), with $p_v \approx 0.224 \times p_v$(MES),
where $p_v$(MES), refers to the full rank maximally entangled state. 
Similar results are obtained for $r=3$ and $d=5$, in which case the state with larger probability of violation corresponds to 
$(\theta_0,\theta_1)\approx(0.840,0.585)$.   For $r=3$ and $d \ge 6$ we did not find any violation. 
\section{Nonlocality of two entangled qudits in the space of behaviours}
In this section we will consider the nonlocality of two entangled qudits in a more general way, by calculating the probability of violation without referring to a
particular Bell inequality. The integration in (\ref{vol}) is now defined in the space of behaviours, characterized by the joint conditional probabilities $\{p(ab|xy)\}$. 
Each $p(ab|xy)$ defines an axis in this space, whose dimension is given by $4d^2$, e.~g., for two inputs and $d$ possible outputs for each of the two parties. This dimension can be lowered if we take into account the normalization of probabilities and the no-signaling condition. With these physical constraints the effective dimension becomes $4d(d-1)$ \cite{Brunner14}. Here we consider qudits from $d=2$ to $d=7$, and the numerical calculations are carried out via linear programming as described in detail in \cite{PhysRevA.82.012118}. The results of this section are summarized in tables \ref{t1} and \ref{t2}. 

\begin{figure}
  \includegraphics[scale=0.5]{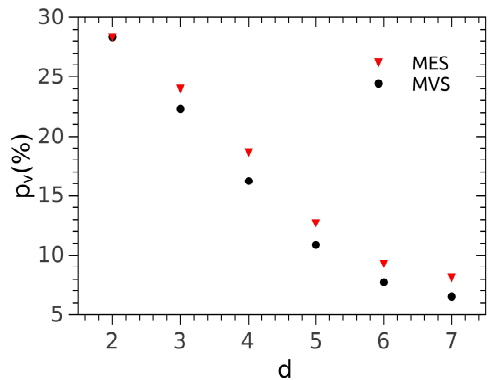}%
 \caption{\label{Fig4}(color online) Probability of violation of MES's (red triangles) and MVS's as functions of the dimension $d$, in the space of behaviours. The nonlocality decreases slowly with the dimension. Note that apart from the qubit case ($d=2$), the MES presents more nonlocality than the MVS. Compare with the monolog plot of Fig. \ref{Fig2}.}
\end{figure}

In accordance with the results of the previous section, the probability of violation decreases as $d$ grows, for 2 observables per party for the investigated values of $d$. However, the fact that there is no restriction to a particular Bell inequality  (all relevant scenarios with a fixed number of observables per party are simultaneously considered), makes the decrease in $p_v$ qualitatively different. Instead of an exponential drop we find an initially linear decay for $2\le d \le5$. In fig. \ref{Fig4} we display the probability of violation (this time in the space of behaviours) for the MES's, red triangles, and, for the sake of comparison, for the MVS considered in the previous section, black squares. Also here, no anomaly shows up.

\begin{figure}
  \includegraphics[scale=0.4]{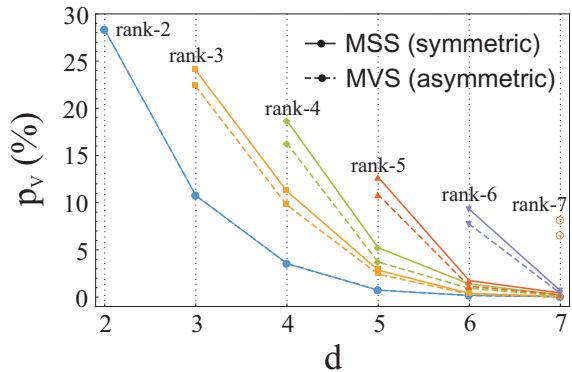}%
 \caption{\label{Fig5}(color online) Probability of violation of maximally symmetric states of several ranks $r$ as a function of $d$, in the space of behaviours. Despite the strong decrease in $p_v$ as $d$ grows,
 all states with $r\ge 2$ present non-vanishing nonlocality. }
\end{figure}

Differently from what we observed in the CGLMP-MBSPS scenario, we found that balanced states with any rank larger than 1, present a nonvanishing probability of 
violation. For instance, with $r=2$ and $d=6$ we found that $0.173\%$ of the possible behaviours are outside the local polytope, while for $r=d=6$
this percentage is about $9.3 \%$. In Fig. \ref{Fig5} we plot $p_v$ against the dimension $d$ for MSS with ranks ranging from $r=2$ to $r=7$.
\begin{table}[h]
\caption{Probability of violation with two measurements settings per party for two qudits MSS and MVS states of different rank. For $r=d$ the MSS's are also MES's (see the bold entries). }
\begin{tabular}{ccccc}
\toprule[1pt]
\multicolumn{2}{c}{} & sample & \multicolumn{2}{c}{$p_v(\%)$} \\
\cmidrule[1pt](l{5pt}r{5pt}){4-5}
$d$ & $r$ & size & $\ket{\psi_{\rm{MSS}}^{\rm{rank}=r}}$ & $\ket{\psi_{\rm{MVS}}^{\rm{rank}=r}}$\\
\midrule[0.75pt]
2 & 2 & $10^{10}$ & \multicolumn{2}{c}{\bf{28.318}} \\\midrule[0.6pt]
3 & 2 & $10^9$ & \multicolumn{2}{c}{10.757} \\
3 & 3 & $10^9$ & \bf{24.011}&22.317\\\midrule[0.6pt]
4 & 2 & $5\times 10^8$ & \multicolumn{2}{c}{3.548} \\
4 & 3 & $5 \times10^8$ & 11.206&9.749\\
4 & 4 & $5\times10^8$ & \bf{18.667}&16.252 \\\midrule[0.6pt]
5 & 2 & $10^8$ & \multicolumn{2}{c}{0.734} \\
5 & 3 & $10^8$ & 2.858&2.423\\
5 & 4 & $10^8$ & 5.228&3.713 \\ 
5 & 5 & $10^8$ & \bf{12.709}&10.863 \\\midrule[0.6pt]
6 & 2 & $10^7$ & \multicolumn{2}{c}{0.173} \\ 
6 & 3 & $10^7$ & 0.397&0.322\\
6 & 4 & $10^7$ & 1.390&0.930 \\
6 & 5 & $10^7$ & 1.748&1.139 \\ 
6 & 6 & $10^7$ &  \bf{9.300}&7.738 \\\midrule[0.6pt]
7 & 2 & $10^6$ & \multicolumn{2}{c}{0.034} \\
7 & 3 & $10^6$ & 0.044&0.029\\
7 & 4 & $10^6$ & 0.215&0.134 \\
7 & 5 & $10^6$ & 0.435&0.258\\
7 & 6 & $10^6$ &  0.679&0.399 \\
7 & 7 & $10^6$ &  \bf{8.132}&6.537\\
\bottomrule[1pt] 
\end{tabular}
\label{t1}
\end{table}
\begin{table}[h]
\caption{Probability of violation for MES with $3\times 3$ measurement settings per party. Compare with the bold entries of
table \ref{t1}.}
	\centering
		\begin{tabular}{c|cccc}
		\toprule[1.1pt]
			d & 2 & 3 & 4 & 5 \\\hline
			$p_v(\ket{\psi_{\rm{MES}}^{\rm{rank}=d}})$ & 78.219 & 78.675 & 71.478 & 56.681 \\\hline
			sample size & $10^9$ & $10^8$ & $10^7$ & $2.25 \times 10^5$ \\
			\bottomrule[1.1pt]
		\end{tabular}
		\label{t2}
\end{table}

Another interesting feature is the strong enhancement in our ability to detect nonlocality by increasing the number of observables per party from 2 to 3 (see table \ref{t2}).
In the simplest case of two entangled qubits, this amounts to a change from $p_v\approx 28.3 \%$ to $p_v \approx 78.2\%$ for MES. 
For $d=r=5$, the probabilities of violation for 2 and 3 observables per party are $12.7\%$ and $56.5\%$, respectively. 
In fact, very recently, this tendency towards large probabilities of violation for an increasing number of observables has been expressed rigorously in \cite{lipinska}.
The property demonstrated in this reference is that, for any pure bipartite entangled state, $p_v$ tends to unity whenever
the number of measurement choices (of the two parties) tends to infinity \cite{lipinska}. 

\begin{figure}
   \includegraphics[width=0.34\textwidth]{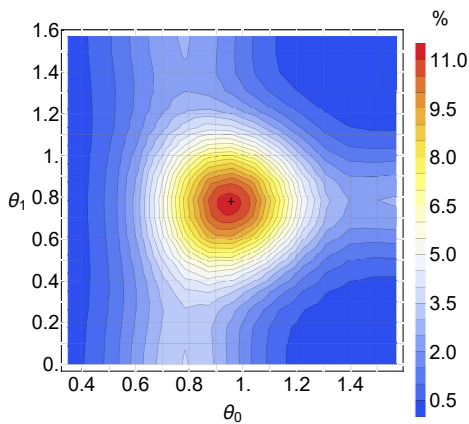}
    \caption{(color online) Probability of violation for rank-3 states with $d=4$ in the space of behavoiurs. The cross corresponds to the state
	$(|00\rangle+|11\rangle+|22\rangle)/\sqrt{3}$. Note how symmetric is this plot in comparison to that of Fig. \ref{G437}.}
\label{fig6}
\end{figure}

Finally, we address the family of states in Eq. (\ref{Par}), this time considering all possible behaviours. The results for the probability of violation are given in the contour plot in Fig. \ref{fig6}. It is much more symmetric than the corresponding contour plot, restricted to the CGLMP-MBSPS scenario, Fig. \ref{G437}. Due to statistical fluctuations, we were not able
to determine the exact location of the state that maximizes the probability, rather, we determined a region in the $\theta_0$-$\theta_1$ plane which contains such a state. The boundary of this region is the innermost contour in Fig. \ref{fig6}, and the MSS with $r=3$ ($d=4$) is identified by the cross.
We thus conclude that the apparent asymmetry revealed in figure \ref{G437} is mainly due to the bias introduced by the choice of a particular facet of the local polytope. Since the number of relevant Bell inequalities grows with the dimension, the effect of this bias tends to increase with $d$.
\section{Closing remarks}
The goal of the present paper was to study quantum nonlocality in bipartite systems of high dimensionality. 
The results showed that the extent of nonlocality decreases with the dimension of the qudits for $d\le 7$ in both, the CGLMP scenario and in the space of behaviours. 
The decay being exponential for the particular Bell inequality we addressed and much slower, at most linear, when all possible behaviours are considered. 
It was additionally shown that, within both approaches, no anomaly of nonlocality showed up, with $p_v$ as the figure of merit. 

The qualitative agreement between the two approaches ceases to hold when maximally symmetric states of lower rank ($r<d$) are considered. While in the fixed Bell scenario
we observed that the MSS are not maximally nonlocal, we found numerical evidence that, whenever the entire local polytope is considered this is no longer true. 
This may be understood as an effect of the increasing (as $d$ grows) bias introduced by the choice of a particular facet. This is a further indication that the probability of violation defined in the space of behaviours is a more fundamental quantity as compared to the volume of violation of a particular Bell scenario.

The regime of large $d$ may be, at least in some sense, considered as a classical limit, and then, we should observe local behaviours as the dominant ones.
However, we may as well conceive the classical limit as a large gathering of two-level systems, which leads to an apparent contradiction. It has been shown 
that the probability of violation strongly increases with the number $N$ of qubits, and two observables per party \cite{guill,us}. In fact, random states of 5 qubits typically present
$p_v> 0.99$ \cite{us} and nonlocality becomes completely dominant for large $N$. 
We remark that this is not a loose comparison because there is an isomorphism between
the Hilbert space of a system with $N$ qubits (for simplicity we assume $N$ to be even) and the Hilbert space of two qudits with $d=2^{N/2}$ levels, each. How do we get opposite trends
in the limit $N \rightarrow \infty$, and consequently in the limit $d \rightarrow \infty$? 

The point is that, in both cases, we have two observables per party, but this amounts to quite different
physical situations. In the $N$-qubit case we have two observables per qubit, say $A_1, A_2; B_1,B_2;C_1,C_2;$ etc. Since each observable is dichotomous, we have 4 possibilities
involving the choice of observables and potential outcomes for every qubit. This leads to a total of $4^N=2^{2N}$ independent possibilities. In the case of 2 qudits with dimension $d=2^{N/2}$ we only have four observables: ${\cal A}_1, {\cal A}_2; {\cal B}_1,{\cal B}_2$, each with $2^{N/2}$ outputs, leading to a total of $4\times 2^{N/2}\times 2^{N/2}=2^{N+2}$ possibilities. So, the four many-output observables in the latter case are not sufficient to compensate for the $2N$ dichotomic observables in the former situation. Of course, in practice, it may become increasingly hard to address individual qubits in the large-$N$ regime.
\acknowledgements
A. F. and F. P. thank the financial support from Conselho Nacional de Desenvolvimento Cient\'{\i}fico e Tecnol\'ogico (CNPq)
and Instituto Nacional de Ci\^encia e Tecnologia-Informa\c{c}\~ao Qu\^antica (INCT-IQ).
A.R and W.L. are supported by the National Science Center (NCN) Grant No. 2014/14/M/ST2/00818.
T.V. is supported by the National Research, Development and Innovation Office NKFIH (Grant Nos. K111734, and KH125096).
\appendix*
\section{CGLMP inequality under multiport beam splitters-phase shifters experimental setup.}
Any probability term $P(A_a=B_b+k)$ in the CGLMP inequality may be written in function of joint probabilities as:
\begin{eqnarray*}
P(A_a=B_b+k)& = &\sum_{j=0}^{d-1}P(A_a=j\oplus k,B_b=j)\\
	     & = & {}\sum_{j=0}^{d-1}P_{ab}(j\oplus k,j),
\end{eqnarray*}
thus, $\mathcal{B}_k$ may be written as: 
$$\mathcal{B}_k=\sum_{a,b=1}^{2}\sum_{j=0}^{d-1}P_{ab}(j\oplus \kappa_{abk},j\oplus \lambda_{abk}), $$
with non vanishing coefficients $\kappa_{abk}$ and $\lambda_{abk}$ given by: $\kappa_{11k}=\kappa_{22k}=\lambda_{12k}=k$, and $\lambda_{21k}=k+1$.

Joint probabilities for the experimental setup considered in this work (equation \ref{PQM}) satisfy the following symmetry property:
\begin{equation*}
\sum_{j=0}^{d-1}P_{ab}(j\oplus k,j\oplus l)=d\cdot P_{ab}(k,l),
\end{equation*}
taking this into account, it is easy to see that $\mathcal{B}_k-\mathcal{B}_{-(k+1)}$ in the CGLMP inequality (eq. \ref{ECGLMP}) reduces to:
\begin{multline*}
 \frac{2}{d}\sum_{a,b=1}^{2}\sum_{m>n}^{d-1} \Re(\alpha_m\alpha^*_{n})\Big\{\cos\Delta\beta^{mn}_{ab}(\kappa_{abk},\lambda_{abk}) \\
-\cos\Delta\beta^{mn}_{ab}(\kappa_{ab(-k-1)},\lambda_{ab(-k-1)})\Big\}.
\end{multline*}

Using trigonometrical identities, the CGLMP function $I_d$ takes the form:
\begin{multline*}
I_d=\sum_{a,b=1}^{2}\sum_{m>n=0}^{d-1}C_{ab}^{mn}\sin\left(\frac{\pi}{d}(m-n)\right)\times\\
\times\Big\{\cos\left(\phi^m_a+\varphi^m_b-\phi^{n}_a-\varphi^{n}_b\right)+\\
+A_{ab}^{mn}\sin\left(\phi^m_a+\varphi^m_b-\phi^{n}_a-\varphi^{n}_b\right)\Big\},
\end{multline*}
with:
\begin{equation*}
A_{ab}^{mn}=(-1)^{a(1+b)+1}\cot\left(\frac{\pi}{d}(m-n)\right)
\end{equation*}
and
\begin{equation}
\label{AmpAP}
C_{ab}^{mn}=\frac{4\Re(\alpha_m\alpha^*_{n})}{d}(-1)^{b(1+a)}\mathcal{C}_{mn},
\end{equation}
where:
\begin{equation*}
\mathcal{C}_{mn}=\sum_{k=0}^{[d/2]-1}\left( 1-\frac{2k}{d-1}\right)\sin\left(\frac{\pi}{d}(m-n)(2k+1)\right).
\end{equation*}

By using the harmonic addition theorem, the CGLMP function for quantum joint probabilities under a measurement scheme based on multiport beam splitters 
and phase shifters characterized by a set of angles $(\varphi^m_b,\phi^{n}_a)$ reduces to:
\begin{equation*}
I_d=\sum_{a,b=1}^{2}\sum_{m>n=0}^{d-1}C_{ab}^{mn}\cos(\phi^m_a+\varphi^m_b-\phi^{n}_a-\varphi^{n}_b+\varPsi_{ab}^{mn}),  
\end{equation*}
with amplitude $C_{ab}^{mn}$ given by \ref{AmpAP} and phase coefficient:
\begin{equation}
\label{PhCGLMP}
\varPsi_{ab}^{mn}= (-1)^{a(1+b)}\left[\frac{\pi}{2}-\frac{\pi}{d}(m-n)\right].
\end{equation}

\bibliography{apssamp.bib}

\end{document}